\documentclass[twocolumn,showpacs,preprintnumbers,amsmath,amssymb,pra]{revtex4}
\usepackage{graphicx}% Include figure files
\usepackage{dcolumn}% Align table columns on decimal point
\usepackage{hyperref}

\begin{document}
\title{Surface plasmon polariton induced strong coherent superposition}

\author{Sintayehu Tesfa$^{1,2}$ and Ali A. Kamli$^1$ }
% \email{sint\_tesfa@yahoo.com}
\affiliation{$^1$Physics Department, Jazan University, P. O. Box 114, Jazan, Saudi Arabia\\
$^2$Physics Department, Addis Ababa University, P. O. Box 1176, Addis Ababa, Ethiopia}

\date{\today}

\begin{abstract}A strong coherent superposition  is shown to be induced between the energy states of a two-level emitter immersed in a surface plasmon polariton evanescently confined to a small metallic ring. It turns out that there is nearly equal chance for the two-level emitter to be found either in the upper or lower energy level at steady state even for a weaker surface plasmon driving strength. This indicates that it is possible to  create a qubit that is robust against emitter spontaneous damping rate, which can be applicable in designing a device that can perform certain quantum information processing tasks. In support of this proposal, the photon statistics of the emitted radiation is found to exhibit significant nonclassical features.
\end{abstract}

 \maketitle

It is a well established fact that when an atom is coupled to an external radiation via the walls of the cavity the structure of the atomic energy levels are considerably altered, wherein it is possible to induce a coherent superposition between the involved energy states \cite{pr1881969,pra414083,pra426873,jmo541759,jmo551587}.  With this background and based on the observed strong analogy between the light and the surface plasmon polariton (SPP),  the chance for inducing the coherent superposition when a two-level emitter interacts with the surface plasmon is explored. Earlier, due to the accompanying tight field confinement and reduced velocity of the surface plasmon, the nanoscale structure is observed to capture the majority of spontaneously emitted radiation into the guided surface plasmon modes \cite{pra79033815,ol371337}. In line with this, embedding an emitter in the SPP was shown to lead to a significant improvement in the stimulated emission profile of the surface plasmon mode \cite{prb70155416,prl94177401,prl101226806}.

In connection with this, the field of plasmonics is currently attracting intense interest mainly in view of the scalability, the strong coherent coupling of the nanoscale structure with the emitters immersed in it, and a remarkable potential it shows to carry out certain quantum information processing tasks \cite{n418304,prl92236801,np3807}. These interests can impart even far-reaching sense  if viewed in terms of unique capabilities, foreseeable applicability, and technical feasibility of the SPP to localize \cite{np483} and manipulate \cite{pr408131} electromagnetic radiation and the emitter \cite{prl109235309}. The technique that enables to instigate strong coherent coupling between individual optical emitters and excitation guided plasmon localized to nanowire has been described in \cite{prl97053002} and  promising candidates for achieving such strong coupling between photons and atoms  have been experimentally realized \cite{n450402,nn5195,np4174}. Moreover, photon-emitter interaction via sub-wavelength confinement of optical fields near metallic nanoscale structure has already been demonstrated \cite{n44539,prl95257403}. In this contribution, owing to the emerging enhanced coupling between the SPP and the emitter placed in it, a mechanism that has a potential to induce a significant coherent superposition between the energy states of the emitter is reported.

 The proposed system mainly comprises a small metallic ring and a two-level emitter placed at a fixed point very close to the outer surface of the ring. The system under consideration is assumed to be pumped by an external coherent light continuously via a tampered waveguide evanescently coupled to the ring. As a result of the interaction of this external pumping coherent light with the electron on the metal,  after quite sometime a strong  surface plasmon mode can be induced on the outer surface of the ring. Once created, the surface plasmon mode is expected to propagate by tracking the geometry of the ring  due to the involved tight confinement. In a  nanowire regime, the higher modes of generated SPP are neglected whereby only the axial modes are assumed to be successfully  involved in the intended interaction \cite{ol371337}.

 The two-level emitter, on the other hand, is assumed to be immersed well in the surface plasmon mode and placed at a fixed position from the surface of the ring in order to neglect the radial dependence of the interaction. In this context, the emitter with small size sees the axially propagating modes which essentially resembles an ordinary external pumping. Since the emitter absorbs one of the surface plasmons at a time, it predominantly emits the radiation in the direction of the propagation of the SPP and hence any potential spontaneous emission in other directions is included in the corresponding spontaneous emitter damping rate.  It is worth noting that nowadays plasmonic devices of this kind and on the assumed scale can be practically realized \cite{jpcc11524469,ol371337}.

Quite obviously the dispersion relation of the propagating surface plasmon modes depends on the involved circular geometry, however, in the nanowire limit, it is possible to treat the metal-air interface as planar. With this consideration, the surface plasmon mode  that propagates at the interface between a metal  and the air is describable by the pertinent wave number that can be expressed for nonmagnetic case  as
\begin{align}k={\omega\over c}\sqrt{{\varepsilon_{m}\over1+\varepsilon_{m}}},\end{align}
  where $\varepsilon_{m}$ is the frequency dependent permittivity of the metal, $\omega$ is the frequency of the SPP, and $c$ is the speed of light. In order to obtain the accompanying dispersion relation, it is preferred to define $\varepsilon_{m}$ in view of the modified Drude model as \cite{cai}
\begin{align}\label{em}\varepsilon_{m}=\varepsilon_{\infty}-{\omega^{2}_{p}\over\omega^{2}+\Gamma^{2}}+i{\omega^{2}_{p}\Gamma\over\omega(\omega^{2}+\Gamma^{2})},\end{align}
where $\Gamma$ is the loss parameter that represents the electron collision rate \cite{prl97053002,prl90027402}, $\omega_{p}$ is the plasma frequency, and $\varepsilon_{\infty}$ is the offset constant that accounts for the contribution from the inter-band transitions of bound electrons in the metal.

 In light of the involved bulk interaction, the loss is often turns out to be a serious challenge. Earlier proposals indicated that this predicament can be circumvented by planting a gain (emitter) in the metal \cite{prl90027402} or by using metamaterial supported by electromagnetically induced transparency of the medium instead of the metal \cite{prl101263601,pra81033839}. However, in this contribution,  the loss of energy due to the absorption of the metal is neglected with the presumption  that the external continuous pumping can somehow compensate for it. Recent numerical calculations show that the parameters that best fit the experimental data of the dispersion relation, for instance, for silver are in the order of $10^{16}s^{-1}$ for $\omega_{p}$ and $10^{13}s^{-1}$ for $\Gamma$ \cite{prl101263601,pra81033839}.
However, for convenience, $\omega_{p}=1$ is used in numerical calculations throughout where the corresponding value of $\Gamma/\omega_{p}$ is taken to be in the region of $10^{-3}$. In the same manner, parameters under consideration including $\omega$, $ck$, and emitter spontaneous damping rate are normalized with respect to $\omega_{p}$.

\begin{figure}[htb]
\centerline{\includegraphics [height=6.5cm,angle=0]{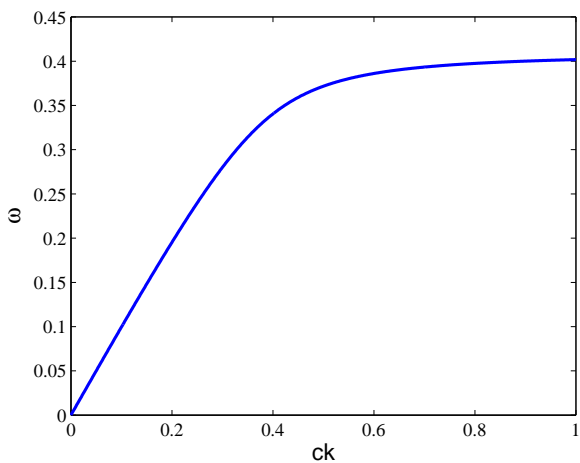}}
\caption {\label{fig1} The real part of the dispersion relation for $\varepsilon_{\infty}=5$ and $\Gamma=10^{-3}$. In numerical calculation,  the loss of energy to the medium via absorption that accounts for the imaginary part of the permittivity is neglected \cite{n424824}. Separate numerical calculation (data not included) shows that the contribution of the imaginary part is quite small for parameters under consideration. Even though two distinct branched bands of frequency can be observed, the one that is lower than the light line, $\omega=ck$ line, is considered.} \end{figure}

 As clearly shown in Fig. \ref{fig1}, the dispersion relation is significantly deviated from the corresponding cavity filled with a vacuum ($\omega=ck$) particularly for larger values of the wave number. For instance, the frequency at $ck=1$ is less by more than half of the corresponding light line.  Generally, such a notable modification in the dispersion relation is responsible for a significant change in the population dynamics which subsequently leads to strong coherent superposition. It is worth noting that a similar form of dispersion relation has also observed earlier \cite{n424824}.

Furthermore, the emitter under consideration can be generally perceived as an electric dipole that interacts with the SPP evanescently confined to the outer surface of the  ring. The pertinent dipole moment is associated with energy levels designated as $|a\rangle$ (excited state) and $|b\rangle$ (lower state) and characterized by an excitation frequency $\omega_{a}=E_{a}-E_{b}$. In line with this consideration, the interaction Hamiltonian that represents a  two-level emitter immersed in a SPP  can be described in the rotating wave approximation by
\begin{align}\hat{H}_{I}&=ig_{s}(\omega_{k})\big[\hat{\sigma}_{+}\hat{a}-\hat{a}^{\dagger}\hat{\sigma}_{-}\big],\end{align} where
$\hat{\sigma}_{+}=|a\rangle\langle b|$ and
$\hat{\sigma}_{-}=|b\rangle\langle a|$
are the usual atomic operators that satisfy the spin-1/2 algebra of the Pauli matrices wherein  $\hat{\sigma}_{+}$ and $\hat{\sigma}_{-}$ describe the excitation and de-excitation of the  emitter, $\hat{a}$ and $\hat{a}^{\dagger}$ are the annihilation and creation operators for SPP.

Moreover, $g_{s}(\omega_{k})$ represents the coupling between the emitter and the surface plasmon mode which is taken to have the form
\begin{align}\label{cc1}g_{s}(\omega_{k})=\omega_{a}|{\vec{\mu}}_{ab}|\sqrt{{1\over2\varepsilon_{0}V\omega_{k}}}{\bf{\hat{e}}}_{k}.{\bf{\hat{u}}}_{d},\end{align} where $|{\vec{\mu}}_{ab}|$ is the dipole moment, $\varepsilon_{0}$ is the permittivity of the free space,  $\omega_{k}$ is the frequency of the surface plasmon mode that depends on the accompanying dispersion relation, ${\bf{\hat{e}}}_{k}$ is the unit vector in the direction of polarization, ${\bf{\hat{u}}}_{d}$ is the unit vector parallel to the direction of the dipole moment, and $V$ is the quantization volume. Treating the ring as some sort of a nanoscale cylindrical shell leads to $g_{s}(\omega_{k})\propto{1\over R}$, where $R$ is the radius of the ring. This implies that the smaller the radius of the ring, the stronger the coupling between the two would be. Despite such a promise, the discussion in this contribution is restricted to the nanowire approximation to limit the otherwise involved mathematical rigor. At this juncture, it may worth emphasizing that the results reported in this contribution can be readily modified for the better by changing the radius and the thickness of the ring which can be taken as motivation for further study of the system under consideration.

The relatively small emitter placed well within the surface plasmon mode experiences a continuous driving in the same way as the atom trapped in the cavity and externally pumped since the azimuthal dependence of the polarization of the SPP can elude it as a result of its size. In this context, driving a two-level emitter on resonance by a surface plasmon mode can be taken as per the pumping of a two-level atom continuously by an external laser beam whose frequency matches the atomic transition frequency. Therefore, upon treating the driving surface plasmon mode classically (setting the operator $\hat{a}$ as a real positive $c$-number $\alpha$), the Hamiltonian that describes the interaction of a two-level emitter with the SPP in the rotating wave and electric dipole approximations can be expressed as
\begin{align}\hat{H}_{I}={i\Omega_{s}\over 2}(\omega_{k})\big[\hat{\sigma}_{+}-\hat{\sigma}_{-}\big],\end{align}
where $\Omega_{s}(\omega_{k})=2g_{s}(\omega_{k})\alpha$ represents the amplitude of the driving surface plasmon mode and can be interpreted in the same way as the Rabi frequency.

The emitter actually recognizes the strength of the pumping mechanism in terms of the intensity of the surface plasmon and the strength of the coupling constant that mainly depends on the size and geometry of the metallic structure. Hence, as one might clearly infer from Fig. \ref{fig1}, the amplitude of the surface plasmon mode that is  perceived by the two-level emitter  decreases with $ck$.  Since the frequency of surface plasmon mode that is supported by the system under consideration is found to be far less than the corresponding light line, the interaction between the emitter and the SPP acquires a far better strength due to the size and geometry of the metallic structure. Quite interestingly, this result indicates that the chance for establishing emitter-SPP entanglement can be far more significant specially for larger values of $ck$ where the accompanying strength of pumping by surface plasmon modes is not so great. Remarkably, this coupling can be enhanced further by reducing the size of the ring. It is hence possible to see in actual practical setting that the coupling can be controlled via the size of the ring and the intensity of the external coherent pumping.

Since the surface plasmon mode is perceived to propagate round-and-round on the surface of the nanoscale ring due its evanescent localization in the radial and lateral directions, the emitter-ring coupled system can be treated as some sort of cavity mechanism \cite{pra79033815}. With this consideration and following the standard procedure \cite{lou}, it is possible to verify  that
\begin{align}{d\hat{\rho}\over dt}&= {\Omega_{s}(\omega_{k})\over2}\big[\hat{\sigma}_{+}\hat{\rho}-\hat{\rho}\hat{\sigma}_{-}-\hat{\sigma}_{-}\hat{\rho}-\hat{\rho}\hat{\sigma}_{-}\big] \notag\\&+{\gamma\over2}\big[2\hat{\sigma}_{-}\hat{\rho}\hat{\sigma}_{+}-\hat{\sigma}_{+}\hat{\sigma}_{-}\hat{\rho} -\hat{\rho}\hat{\sigma}_{-}\hat{\sigma}_{+}\big],\end{align} where $\rho$ is the density operator that describes the population dynamics of the emitter and $\gamma$ is the spontaneous damping rate of the emitter. It is a well established fact that the emitter spontaneous damping rate is also different from the corresponding cavity scenario. In this contribution, for the sake of convenience, $\gamma$ is taken to be constant but with larger values than actually expected.

Making use of the master equation, the time evolution of the expectation values of the atomic creation, atomic annihilation, and energy operators (defined as
$\hat{\sigma}_{z}=|a\rangle\langle a|-|b\rangle\langle b|$) are found to be
\begin{align}{d\over dt}\langle\hat{\sigma}_{-}(t)\rangle=-{\gamma\over2}\langle\hat{\sigma}_{-}(t)\rangle-{\Omega_{s}(\omega_{k})\over2}\langle\hat{\sigma}_{z}(t)\rangle,\end{align}
\begin{align}{d\over dt}\langle\hat{\sigma}_{+}(t)\rangle=-{\gamma\over2}\langle\hat{\sigma}_{+}(t)\rangle-{\Omega_{s}(\omega_{k})\over2}\langle\hat{\sigma}_{z}(t)\rangle,\end{align}
\begin{align}{d\over dt}\langle\hat{\sigma}_{z}(t)\rangle&=-\gamma\big(\langle\hat{\sigma}_{+}(t)\rangle+1\big)\notag\\&+\Omega_{s}(\omega_{k})\big[\langle\hat{\sigma}_{-}(t)\rangle +\langle\hat{\sigma}_{+}(t)\rangle\big].\end{align}

Upon solving these coupled differential equations, it is possible to study the population dynamics of the emitter and the property of the radiation. For instance, the population dynamics of the two-level emitter can be described by the probability of the emitter to be in the upper energy level, which can be expressed as
\begin{align}\rho_{aa}(t)={1\over2}\big[\langle\hat{\sigma}_{z}(t)\rangle+1\big].\end{align} Following the method outlined in \cite{jmo541759}, this expression is found to take at steady state the form
\begin{align}\rho_{aa}={\Omega_{s}(\omega_{k})^{2}\over2\Omega_{s}(\omega_{k})^{2}+\gamma^{2}}.\end{align}

\begin{figure}[htb]
\centerline{\includegraphics [height=6.5cm,angle=0]{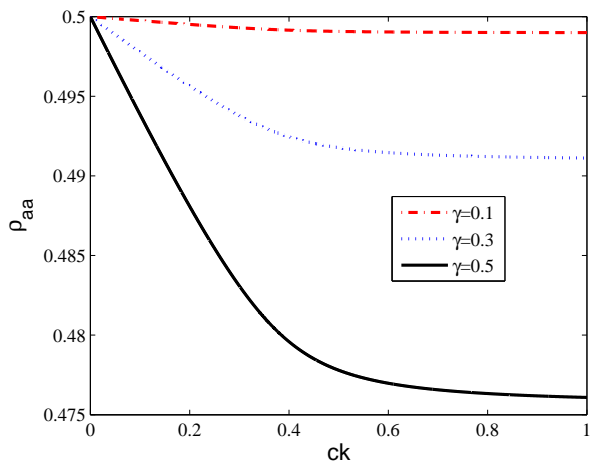}}
\caption {\label{fig4} The probability for the emitter to be in the upper energy level at steady state for $\Gamma=10^{-3}$, $\varepsilon_{\infty}=5$, and different values of $\gamma$. Even though the cause of the spontaneous damping of the emitter to the environment originates from various physical processes, in the scale of the system under consideration, the spontaneous emission rate to the surface plasmon mode is very significant \cite{pra69013812} which implies that $\gamma$ is actually smaller than what is considered here.  Following similar approach, it was shown earlier that $\rho_{aa}\approx0.5$ for a strong and $\rho_{aa}\approx0$ for a weak external driving when the atom is assumed to be trapped in a resonant cavity \cite{jmo541759,jmo551587}.}\end{figure}

As one can clearly see from Fig. \ref{fig4},   the probability for the emitter to be in the upper energy level at steady state is very close to $50\%$. This generally indicates that the surface plasmon can drive the emitter to the upper energy level successfully in the same manner as the corresponding two-level atom is driven with a strong external coherent radiation in the cavity scenario. Contrary to earlier observation of nearly $50\%$ probability of finding the atom trapped in a cavity in the upper energy level when driven externally by a strong radiation \cite{jmo541759,jmo551587}, critical scrutiny of Fig. \ref{fig4} reveals that it is possible to pump the emitter to the upper energy level efficiently especially when the spontaneous decay rate is comparatively smaller without necessarily requiring intense surface plasmon mode. This outcome can be taken as encouraging result for creating a robust qubit from the energy states of the emitter by making use of the coupling associated with the nanoscale confinement. It is worth noting that the qubit to be created is still quite sensitive to the corresponding spontaneous damping rate.

Furthermore, it is a well established fact that the photon statistics of the emitted radiation can
be studied by using the normalized two-time second-order correlation function that can
be expressed in terms of the excitation and de-excitation atomic operators in the form
\begin{align}g^{(2)}(\tau)={\langle\hat{\sigma}_{+}(t)\hat{\sigma}_{+}(t+\tau)\hat{\sigma}_{-}(t+\tau)\hat{\sigma}_{-}(t)\rangle\over \langle\hat{\sigma}_{+}(t)\hat{\sigma}_{-}(t)\rangle^{2}}.\end{align}
Following the procedure developed for cavity scenario, the corresponding normalized two-time second-order correlation function  in the strong driving limit and at steady state is found to reduce to
\begin{align}g^{(2)}(\tau)=1-\cos(\Omega_{s}(\omega_{k})\tau)e^{-3\gamma\tau/4}.\end{align}

\begin{figure}[htb]
\centerline{\includegraphics [height=6.5cm,angle=0]{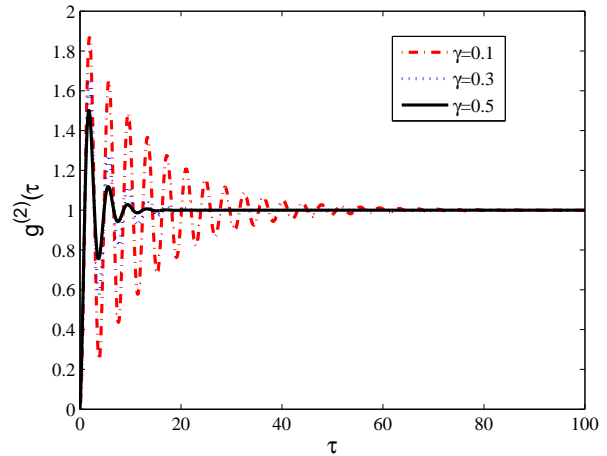}}
\caption {\label{fig8} Plots of the two-time correlation function at steady state versus the delayed time for $ck=0.5$, $\Gamma=10^{-3}$, $\varepsilon_{\infty}=5$, and different values of $\gamma$. The nonclassical features of the emitted radiation would be significant for shorter time of delayed coincidence. Quite interestingly, the nonclassical features exists even for larger values of $\gamma$.} \end{figure}

The anti-bunching phenomenon and the sub-Poissonian character shown in Fig. \ref{fig8} indicate that it is possible to ascribe a quantum content to the emitted radiation at least under certain specific conditions. Since the surface plasmon mode can induce a coherent superposition between the upper and lower energy states of the emitter,  the manifestation of the nonclassical feature can be linked to this induced atomic coherent superposition. With this background, it would not be hard to envisage that if more number of emitters are embedded in the surface plasmon, there can be a real chance of entangling the involved atomic states via the exchange of the available radiation \cite{ol371337}. Such an understanding, on the other hand, can be taken as a starting point for looking into the possibility of developing similar mechanisms for carrying out realizable quantum information processing tasks.

In conclusion, the population dynamics of the emitter which is relevant in inducing the coherent superposition of the atomic states of the emitter is found to be significantly modified. It is also clearly shown that the speed of the surface plasmon mode supported by the system under consideration can be made far less than the corresponding cavity situation under appropriate selection of the parameters. Such a possibility of observing reduced speed implies that the emitter can enjoy a relatively longer time to interact with the surface plasmon mode that can be responsible for establishing a stronger coupling between the emitter and SPP.  In this context,  even when a substantial damping rate due to spontaneous emission out of the prescribed states are considered, the chance for observing nearly $50\%$ of the emitter to be in the upper energy level, which can be taken as a basis for creating a robust qubit from the energy states of the emitter, is found to be quite significant. In addition, in relation to the assumed circular geometry and continuous external pumping, it is possible to induce intense surface plasmon mode wherein the photon statistics of the emitted radiation can reveal nonclassical features for certain parameters. The quantum content of the radiation can be linked predominantly to the coherent superposition that can be induced by the surface plasmon mode. Hence based on the observed significant modification in coupling between the emitter and SPP, it is expected that reliable and scalable device for carrying out certain quantum information processing tasks can be designed from such nanoscale structure.


\begin{thebibliography}{1}
%%%%
\bibitem{pra414083} R. D'Souza, A. S. Jayarao, and S. V. Lawande, Phys. Rev. A {\bf{41}}, 4083 (1990).
\bibitem{jmo541759} S. Tesfa, J. Mod. Opt. {\bf{54}}, 1759 (2007).
\bibitem{pr1881969} B. R. Mollow, Phys. Rev. {\bf{188}}, 1969 (1969).
\bibitem{pra426873} A. S. Parkins, Phys. Rev. A {\bf{42}}, 6873 (1990).
\bibitem{jmo551587} S. Tesfa, J. Mod. Opt. {\bf{55}}, 1587 (2008).
\bibitem{ol371337} G. Y. Chen, C. M. Li, and Y. N. Chen, Opt. Lett. {\bf{37}}, 1337 (2012).
\bibitem{pra79033815} Y. N. Chen, G. Y. Chen, D. S. Chuu, and T. Brandes, Phys. Rev. A {\bf{79}}, 033815 (2009).
\bibitem{prb70155416} I. Avrutsky, Phys. Rev. B {\bf{70}}, 155416 (2004).
\bibitem{prl101226806} M. A. Noginov, G. Zhu, M. Mayy, B. A. Ritzo, N. Noginova, and V. A. Podolskiy, Phys. Rev. Lett. {\bf{101}}, 226806 (2008).
\bibitem{prl94177401} J. Seidel, S. Grafstr\"{o}m, and L. Eng, Phys. Rev. Lett. {\bf{94}}, 177401 (2005).
\bibitem{n418304} E. Altewischer, M. P. van Exter, and J. P. Woerdman, Nature {\bf{418}}, 304 (2002).
\bibitem{prl92236801}  E. Moreno, F. J. G. Vidal, D. Erni, J. I. Cirac, and L. M. Moreno, Phys. Rev. Lett. {\bf{92}}, 236801 (2004).
\bibitem{np3807}  D. E. Chang, A. S. S{\o}rensen, E. A. Demler, and M. D. Lukin, Nature Phys. {\bf{3}}, 807 (2007).
\bibitem{np483} D. K. Gramotnev and S. I. Bozhevolnyi, Nat. Photonics {\bf{4}}, 83 (2010).
\bibitem{pr408131} A. V. Zayats, I. I. Smolyaninovb, and A. A. Maradudin, Phys. Rep. {\bf{408}}, 131 (2005).
\bibitem{prl109235309} M. Gullans, T. G. Tiecke, D. E. Chang, J. Feist, J. D. Thompson, J. I. Cirac, P. Zoller, and M. D. Lukin, Phys. Rev. Lett. {\bf{109}}, 235309 (2012).
\bibitem{prl97053002} D. E. Chang, A. S. S{\o}rensen, P. R. Hemmer, and M. D. Lukin, Phys. Rev. Lett. {\bf{97}}, 053002 (2006).
\bibitem{nn5195} T. M. Babinec, B. J. M. Hausmann, M. Khan, Y. Zhang, J. R. Maze, P. R. Hemmer, and M. Loncar, Nature Nanotech. {\bf{5}}, 195 (2010).
\bibitem{np4174} J. Claudon, J. Bleuse, N. S. Malik, M. Bazin, P. Jaffrennou, N. Gregersen, C. Sauvan, P. Lalanne, and J. M. G\`{e}rard, Nat. Photonics {\bf{4}}, 174 (2010).
\bibitem{n450402} A. V. Akimov, A. Mukherjee, C. L. Yu, D. E. Chang, A. S. Zibrov, P. R. Hemmer, H. Park, and M. D. Lukin, Nature {\bf{450}}, 402 (2007).
\bibitem{n44539} C. Genet and  T. W. Ebbesen, Nature {\bf{445}}, 39 (2007).
\bibitem{prl95257403} H. Ditlbacher, A. Hohenau, D. Wagner, U. Kreibig, M. Rogers, F. Hofer, F. R. Aussenegg, and J. R. Krenn, Phys. Rev. Lett. {\bf{95}}, 257403 (2005).
\bibitem{jpcc11524469} S. D. Liu, Z. Yang, R. P. Liu, and X. Y. Li, J. Phys. Chem. C {\bf{115}}, 24469 (2011).
\bibitem{cai} W. Cai and V. Shalaev, {\it{Optical metamaterials: Fundamental and applications}} (Springer Science+Business Media, New York, 2010).
\bibitem{prl90027402} D. J. Bergman and M. I. Stockman, Phys. Rev. Lett. {\bf{90}}, 027402 (2003).
\bibitem{prl101263601} A. Kamli, S. A. Moiseev, and B. C. Sanders, Phys. Rev. Lett. {\bf{101}}, 263601 (2008).
\bibitem{pra81033839} S. A. Moiseev, A. A. Kamli, and B. C. Sanders, Phys. Rev. A {\bf{81}}, 033839 (2010).
\bibitem{n424824} W. L. Barnes, A. Dereux, and T. W. Ebbesen, Nature {\bf{424}}, 824 (2003).
\bibitem{lou} W. H. Louisell,  {\it{Quantum Statistical Properties of Radiation}} (Wiley, New York, 1973).
\bibitem{pra69013812} V. V. Klimov and M. Ducloy, Phys. Rev. A {\bf{69}}, 013812 (2004); D. E. Chang, A. S. S{\o}rensen, P. R. Hemmer, and M. D. Lukin, Phys. Rev. A {\bf{76}}, 035420 (2007).
\end{thebibliography}
\end{document}